# Unsupervised Speaker Diarization in Distributed IoT Networks Using Federated Learning

Amit Kumar Bhuyan, Hrishikesh Dutta, and Subir Biswas

***Abstract* – This paper presents a computationally efficient and distributed speaker diarization framework for networked IoT-style audio devices. The work proposes a Federated Learning model which can identify the participants in a conversation without the requirement of a large audio database for training. An unsupervised online update mechanism is proposed for the Federated Learning model which depends on cosine similarity of speaker embeddings. Moreover, the proposed diarization system solves the problem of speaker change detection via. unsupervised segmentation techniques using Hotelling's t-squared Statistic and Bayesian Information Criterion. In this new approach, speaker change detection is biased around detected quasi-silences, which reduces the severity of the trade-off between the missed detection and false detection rates. Additionally, the computational overhead due to frame-by-frame identification of speakers is reduced via. unsupervised clustering of speech segments. The results demonstrate the effectiveness of the proposed training method in the presence of non-IID speech data. It also shows a considerable improvement in the reduction of false and missed detection at the segmentation stage, while reducing the computational overhead. Improved accuracy and reduced computational cost makes the mechanism suitable for real-time speaker diarization across a distributed IoT audio network.**

***Keywords* – Unsupervised Learning, Bayesian Methods, Federated Learning, Distributed Processing, Hotelling's t-squared statistic, Bayesian Information Criterion, Cepstral Analysis**

*Note to the Editor and Reviewers: We have addressed all the comments and concerns raised by the Associate Editor and the reviewers. Detailed responses to each comment and the corresponding modifications made to the manuscript are provided in the accompanying rebuttal document.*

Nomenclature

| Symbol | Definition |
|---|---|
| $MFCC$ | Mel Frequency Cepstral Coefficients |
| $BIC$ | Bayesian Information Criterion |
| $S$ | General Notation for Window |
| $N_S$ | General Notation for Window size |
| $X$ and $Y$ | General Notation for Sub-windows |
| $N_X$ and $N_Y$ | General Notation for Sub-window sizes |
| $\vartheta_s$ | Distribution parameter for $S$ |
| $L_0$ | Log likelihood of $\vartheta s$ given $S_i \in S$ |
| $\vartheta_x$ and $\vartheta_y$ | Distribution parameter for $X$ and $Y$ |
| $L_1$ | Log likelihood of $\vartheta_x$ and $\vartheta_y$ |
| $\Delta BIC$ | Change in Bayesian Information |
| $\Delta K$ | Difference in the number of parameters |
| $\lambda$ | Penalty factor for model fit vs. complexity |
| $Q_S$ | Quasi-Silences from Spectral Subtraction |
| $A_W$ | Analysis Window around a quasi-silence |
| $N_{aw}$ | Size of Analysis Window |
| $W$ | Window inside an Analysis Window |
| $w1$ and $w2$ | Sub-windows inside window $W$ |
| $w_s$ | Window stride to compute sub-windows |
| $N_{slid}$ | Window slide size inside the $A_W$ |
| $N_g$ | Window grow size inside the $A_W$ |
| $T^2$ | Hotelling's $t^2$-statistic |
| $\overline{\mu_x}$ and $\overline{\mu_y}$ | Sample means of the Sub-windows |
| $\delta_{x,y}(\bar{\mu})$ | Difference between $\overline{\mu_x}$ and $\overline{\mu_y}$ |
| $\Sigma$ | Covariance matrix of the window $S$ |
| $\theta_0$ | Model Weight distribution |
| $E$ | Client Rounds |
| $\theta_t^k$ / $\Delta L_k(\theta_t)$ | Client Model Updates |
| $\theta_{t+1}$ | Server Aggregation |
| $\mu\lvert\lvert\theta - \theta_t\rvert\rvert^2$ | Proximal Term |
| $\mu$ | Regularization parameter |
| $v_t$ | Momentum |
| $\beta$ | Momentum coefficient |
| $\eta$ | Learning Rate |
| $\gamma$ | Exponential Decay Rate |
| $T$ | Communication Rounds |
| $f_{stragglers}$ | Straggler Fraction |
| $N$ or $G$ | Total Samples or per group |
| $L_e$ | Size of Embedding Layer |
| $D_C$ | Clusters from the Clustering stage |
| $D_T$ | Training Audio Data |
| $CD$ | Cluster Segment Embedding |
| $TD$ | Embeddings of training audio data $D_T$ |
| $S_C$ | Cosine Similarity between $CD$ and $TD$ |
| $FDR$ | False Detection Rate for change points |
| $MDR$ | Missed Detection Rate for change points |
| $F_{seg}$ | Accuracy of speaker change detection |
| $Purity\%$ | Correct out of total detected change points |
| $Coverage\%$ | Correct out of total true change points |
| $FAR$ | False Acceptance Rate for speaker ID |
| $FRR$ | Ratio of speakers that were falsely rejected |
| $F_{ID}$ | Accuracy in speaker identification |

The authors are with Electrical and Computer Engineering Department of Michigan State University, East Lansing, USA, 48823. (Email: *bhuyanam@msu.edu, duttahr1@msu.edu, sbiswas@msu.edu*)

## I. Introduction

*Speaker Diarization* is the process [1] of partitioning a conversation-generated audio stream into segments according to the speaker identities. When used together with automatic speaker identification (ASI) systems, by providing the speaker's true identity, diarization can be used to answer the question "who spoke when?" With recent proliferation of ubiquitous and intelligent mobile devices, remote work, meetings, medical diagnosis, and explosive deployment of conversational AI assistants such as Amazon Echo and Apple Siri, the demand for speaker diarization has skyrocketed. It is an essential component for a variety of applications such as call center services, meeting transcriptions, etc [1-6]. In this paper, we develop the components of diarization by delving into different types of segmentation methods, and using distributed Federated Learning (FL) across multiple recording devices. One of the major challenges for diarization is that its performance depends heavily on application-specific constraints. For example, diarization focused on call center audio is mostly about separating just two speakers, often in quite diverse acoustic environments. Diarization for meeting audio, on the other hand, has to deal with multiple speakers. A robust design needs to handle such diverse scenarios.

Many diarization systems in the literature [8-15, 23] suffer from implementation limitations in that an automatic speaker identification (ASI) system must be trained *a priori*, which requires a large speech database. In cases where speech samples for the participants in a conversation are not known beforehand, such methods do not work for on-the-fly operation. Additionally, speaker-specific audio data acquisition is more likely to be conducted in standalone, portable or handheld devices. However, uninterrupted accessibility of audio data from disjointed devices by forming an interconnected network is limited in real-time due to the unrestricted mobility of such handheld devices. Furthermore, diarization systems such as the ones proposed in [32-41] rely solely on the ASI for speaker segmentation on a frame-by-frame basis. This adds redundant decision-making as the continuous segments belonging to one speaker must be determined using decisions from every frame. This makes it computationally heavy and not suitable for embedded platforms. Another class of mechanisms, namely, distance-based segmentation and diarization such as WinGrow [7-15], can deduce the expected result without prior information about the conversation structure or participants, and can be computationally feasible for embedded systems. However, the accuracy of those mechanisms is usually limited by the type and effectiveness of the statistical measures used to calculate the difference between speaker segments. These algorithms are also prone to false speaker change detections due to factors such as pauses, spikes, audio disruptions, overlapped speech, and other statistical anomalies, which are not immediately detectable from audio features. Additionally, in order to determine speaker change points, most distance-based segmentation algorithms use greedy methods [7-9] which are computationally intensive.

To address these shortcomings, this paper proposes a decentralized FL-based speaker diarization mechanism that uses unsupervised segmentation and FL-based speaker identification. The goal is to answer the question of "Who spoke when?" in an unforeseen environment. It focuses on diarization in a distributed scenario in which an array of networked devices (i.e., IoTs) record audio from speakers in a conversation, train for diarization, and share the trained models amongst themselves for improving network-wide diarization performance. Note that each of those devices may not have access to audio from all participants in a conversation. Individually, they may have access to only a part of the conversation (i.e., typically from one or a subset of speakers) based on which the local on-device learning needs to take place. The other distinctiveness of the targeted environment is that the environment itself may be unknown in that the diarization system may not have been trained with *a priori* audio samples from the participating speakers. This renders much-used supervised learning-based techniques largely ineffective. The proposed diarization mechanism deals with such unknown distributed processing environments using unsupervised segmentation and FL.

The proposed mechanism is applied on a speaker diarization system which is a combination of three integral components, namely, speaker segmentation, speaker clustering and speaker identification. The first aims at finding speaker change points in an audio stream. The second aims at clustering or grouping together speech segments on the basis of speakers' acoustic characteristics. The third associates a speaker's identity with the grouped audio segments. The goal is to segment the audio and identify the source of the segmented audio with high accuracy. This goal is achieved by employing unsupervised segmentation and FL-based speaker identification.

Specific contributions of the paper are as follows. First, a FL-based speaker diarization mechanism for distributed audio-recording devices/IoTs is proposed. Second, a novel client device grouping method, namely Random Client Grouping ($RCG$), is introduced for federated model aggregation. This method utilizes the vicinity of interconnected devices and the respective audio data accessibility as a fundamental concept for model aggregation. Third, unsupervised distance-based Bayesian methods, namely, Bayesian Information Criterion ($BIC$) and Hotelling's t-squared statistic ($t^2$-*statistic*), are employed for speaker segmentation and clustering. The advantages of using $t^2$-*statistic* as compared to other statistical methods in terms of segmentation accuracy and computational rigor is analyzed, leading to a hybrid segmentation technique involving both $t^2$-*statistic* and $BIC$. Fourth, a novel unsupervised online update method for FL model is employed based on cosine similarity of speaker embeddings. Finally, the proposed framework is functionally verified and experimentally evaluated with real-world audio conversations from zoom meetings and online sources including podcasts, YouTube, etc. It is demonstrated that the proposed system can achieve performance comparable to centrally trained models, and that is in the absence of *IID* audio data availability and *a priori* training at the audio recording IoT devices.

## II. Related Work

Speaker segmentation is a crucial aspect of diarization systems and has garnered considerable attention from researchers. Existing literature on segmentation often focuses on distance or metric-based approaches, utilizing accuracy as the primary performance metric, derived from false detection and missed detection rates. In contrast to WinGrow's [11] approach, a top-down method called divide-and-conquer (DACDec) [8] was introduced in a previous study, emphasizing distance-based segmentation. However, DACDec's sequence of change point detection raises concerns, as a false detection can impact the identification of temporally proximate true change points. This work addresses these concerns by ensuring that a false detection does not affect subsequent change detection, concentrating the search for the next change point around quasi-silence, regardless of the previous change point's location.

Pitch, as suggested by authors in [12] and [13], replaces Cepstral analysis methods like Mel Frequency Cepstral Coefficient (*MFCC*) in certain instances. The uniqueness of pitch across individuals, its perceivability using Gamma correction function, and its ability to represent a signal without a distribution motivated its use in [12]. Meanwhile, [13] employed pitch estimation via Kalman Filters for speaker change point detection, introducing computational complexity. However, the use of Gamma correction function in this context does not effectively mitigate false alarms. In this paper's approach, quasi-silence-based segmentation addresses the threshold selection issue by computing multiple $\Delta BIC$ values for one window and selecting the highest value as the change point. Furthermore, the proposed method evaluates change points at quasi-silences, eliminating the need for a continuous recursive search.

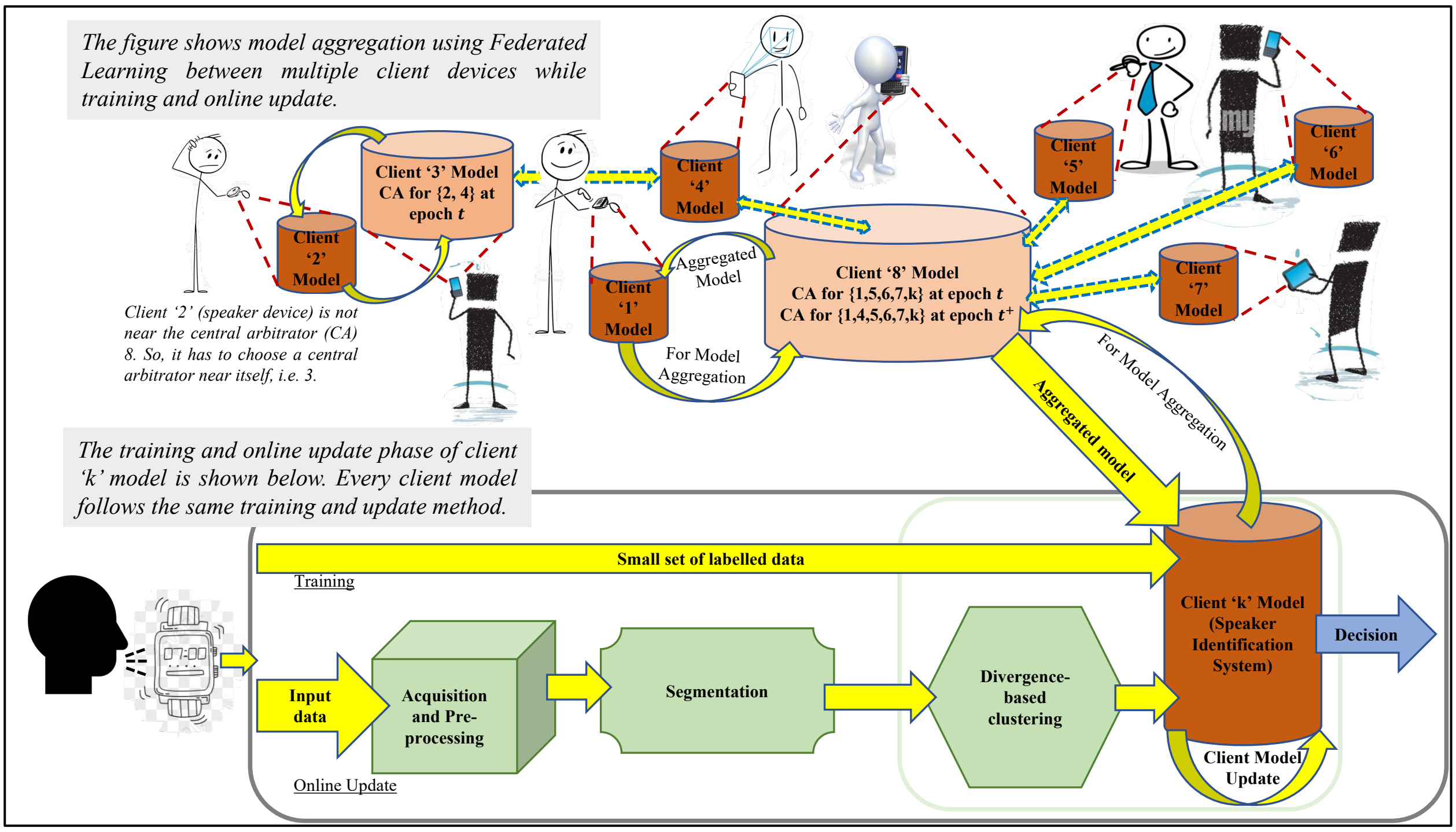

Fig. 1. Federated Learning-Based Distributed Speaker Diarization System

Previous studies [14] and [9] have reported the use of Mahalanobis and cosine-based distances for speaker segmentation. While [14] achieved increased segmentation accuracy at the expense of additional clustering, [9] observed a gradual increase in missed detection with decreased false detection. Our focus is on minimizing the drastic tradeoff between false detection and missed detection in our work, aiming to preserve overall segmentation accuracy.

Researchers have explored features beyond short-term ones like *MFCC* for speaker segmentation. [15] utilized long-term features such as jitter, shimmer, glottal-to-noise (GNE), pitch, and formant to enhance segmentation accuracy. However, the computational complexity of $BIC$ calculation from joint likelihood and long-term features computation limits the applicability of such methods.

In diarization process clustering, [23] proposed an improvement but assumed known numbers of speakers and time segmentations, making the segmentation impractical. In speech separation, various methods handle mixtures with a variable number of speakers. Iterative "one-vs-rest" approaches like [24-27] increase complexity linearly with the number of speakers. Attractor-based strategies like Deep Attractor Network (DANet) [28] require prior knowledge of the speaker count. Anchored DANet [29] mitigates challenges but has limited scalability. Efforts have been made to devise end-to-end approaches, like the Set Transformer ([30]) and end-to-end clustering ([31]), but these come with predefined output constraints.

Other approaches involve inserting speaker role tags into transcripts ([32]) and employing Speaker-Attributed ASR (SA-ASR) ([33-37]), which includes a turn detection mechanism. Target Speaker ASR (TS-ASR) ([38-41]) is designed for diarizing target speaker speech. However, these approaches are limited in scalability, computational intensity, and dependency on specific profiles.

Owing to the requirement of a distributed algorithm to fit the scenario this work is trying to accomplish, federated learning has been explored. Although there is no significant literature available in the application of federated learning for speaker diarization, substantial work has been done to accomplish sparse training [42], an attribute that can be leveraged for this work. *Non-IID* data that results from heterogeneity in data distribution is another attribute that has been explored in works like [44, 45] leading to robustness in model trained using federated learning.

From the literature above, it can be observed that the existing frameworks for speaker diarization generally have four major drawbacks. First, accuracy improvement comes from features that are computationally heavy. Second, the high overall accuracies of the proposed methods come with often unacceptable missed detection rates. Third, the search for change point is continuous which is computationally intensive. Fourth, to determine the identity of a speaker involved in the conversation, intensive training is required. This paper addresses all of the limitations via. an Unsupervised FL Based Speaker Diarization system. The objective is to develop segmentation mechanisms that balances between overall accuracy, acceptable miss detection rates, and manageable computational complexity that is conducive for potential IoT-style audio networks. Additionally, the system should update without external supervision, therefore improving speaker identification accuracy.

## III. System Architecture

The proposed Federated Learning-based [19] speaker diarization framework involves networked audio recording devices, called *client devices*, with limited data processing capabilities. Each device records audio from nearby speakers only and trains a local model using its sparse data. These models are shared with neighboring devices for federated training. The final model assigns speaker identities to audio segments using unsupervised segmentation and clustering techniques, as shown in Fig. 1.

After recording a conversation, each client device processes the audio to extract low-dimensional features that capture its acoustic characteristics. These features are fed into the *Segmentation* stage, where the conversation is divided into segments based on statistical divergence between consecutive audio segments. This segmentation uses a hybrid of the $BIC$ and $t^2$-*statistic*. The segmented audio is then clustered into groups of acoustically similar segments in the *Clustering* stage, assuming each cluster represents a specific speaker. The FL-based *Identification* system assigns identities to these clusters, with its efficiency depending on training state, initial data volume, and online model updates the details of which are discussed in Section V.

Unlike the other existing technologies, the proposed mechanism is able to perform diarization in the presence of limited availability of the recorded audio data. It also has a relatively lighter computational complexity, which is achieved by distributing the role of a single FL aggregator/arbitrator to many client devices. Fig. 1 shows that at epoch $t$, *client device 8* acts as the central arbitrator (CA) for its neighborhood which consists of client devices 1, 5, 6, 7 and *k*. Similarly, at epoch $t$ *client device 3* acts as a central arbitrator for devices 2 and 4. Client 4 migrates towards CA 8 at epoch $t^+$. Detailed explanation of online model update and aggregation mechanisms are presented in Section V.

Before getting into the details of distributed FL, it is important to understand the segmentation and clustering mechanisms, which are explain in the next section.

## IV. Unsupervised Speech Segmentation and Clustering

Before segmenting a conversation, it needs to be pre-processed which involves feature extraction that captures the acoustic characteristics of the conversation. Mel Frequency Cepstral Coefficients (*MFCC*) [16] is a cepstral analysis used for such processing. Every audio frame is first represented as a vector of 12 *MFCC* coefficients, the dimension of which has been empirically chosen [23]. These 12-D vectors are the inputs to the speaker segmentation stage where statistical divergence (distance) is computed for speaker change point calculation. This section explains the employed methods for statistical divergence, segmentation, and clustering.

### *A. Bayesian Information Criteria for Statistical Divergence*

Separating consecutive audio segments is modeled using $BIC$ [8, 11, 23], computed for windows of *MFCC* vectors representing a 12-D distribution. The divergence, $\Delta BIC$, determining speaker change points is computed as follows:

$$L_0 = \sum_{i=1}^{N_S} log\, p(S_i | \vartheta_S) \qquad (1)$$

$$L_1 = \sum_{i=1}^{N_x} log\, p(X_i | \vartheta_x) + \sum_{i=1}^{N_y} log\, p(Y_i | \vartheta_y) \qquad (2)$$

where $N_s = N_x + N_y$, $S$ is the window, and $X$ and $Y$ are sub-windows. $L_0$ is the log likelihood of $\vartheta_s$ given $S_i \in S$ , and $L_1$ of $\vartheta_x$ and $\vartheta_y$ given $X_i \in X$ and $Y_i \in Y$. Bayesian information divergence is then calculated as:

$$\Delta BIC = L_1 - L_0 - \frac{\lambda}{2}. \Delta K. logN_S \qquad (3)$$

In other words, the problem can be stated as follows:

$$\max_{\vartheta_s, \vartheta_x, \vartheta_y} \left[ L_1 - L_0 - \frac{\lambda}{2}. \Delta K. logN_S \right] \qquad (3a)$$

s.t.

$$\sum_{i=1}^{N_S} log\, p(S_i | \vartheta_S) < \sum_{i=1}^{N_x} log\, p(X_i | \vartheta_x) + \sum_{i=1}^{N_y} log\, p(Y_i | \vartheta_y)$$

$$\forall S = 1, 2, 3, \cdots\cdots \qquad (3b)$$

Here, $\lambda$ is a penalty factor [23], and $\Delta K$ is the difference in parameters used for $L_0$ and $L_1$. The default $\lambda$ value is 1. The method uses highest positive $\Delta BIC$ to determine change points where sub-windows belong to different speakers.

### *B. Quasi-Silence based Speaker Segmentation using BIC*

The proposed segmentation method calculates change points around quasi-silences [23], which include both speaker-switching and non-switching pauses. This reduces the number of divergence calculations and increases the likelihood of detecting change points. Quasi-silences are identified using spectral subtraction [23], and their surrounding non-silent regions are analyzed. An analysis window is set around these quasi-silences for applying the *BIC*-based segmentation technique.

The procedure of the proposed method is summarized below.

**Algorithm 1:** Quasi-Silence Based Segmentation using *BIC.*

**Input:** Signal '$S$' and list of quasi-silences '$Qs$'
**Output:** Segment boundaries '$I$'
**for** $j = 0$ to $length(Qs)$ **do**
  $A_W = S(1, N_{aw})$ // *Analysis window around Qs*
  $W = S(1, N_w)$ // *Window of size $N_w$ inside $A_W$*
  $(\Delta BIC_{max}, i_{max}) \longleftarrow$ compute $\Delta BIC$
  **if** $N_w < N_{aw}$ **then do**
    **if** $\Delta BIC_{max} > 0$ **then do**
      $i_{change} = i_{max}$
      Slide Window by $N_{slid}$
    **else**
      Grow Window by $N_g$
    **end if**
  **end if**
  $I \longleftarrow i_{change}$
**end for**

In this segmentation method, a window $W$ is selected within an analysis window $A_W$. Each $W$ is divided into sub-windows $w1$ and $w2$. The size of the sub-windows depends on the stride size $w_s$, with varying strides splitting the window into different sub-window pairs. Each sub-window generates a sample distribution for BIC calculations. The change in $BIC$ ($\Delta BIC$) between sub-windows determines the presence of a change point. If a change point is found, the window slides within $A_W$ by $N_{slid}$, limited by $N_{aw}$. If no change is detected, the window grows by $N_g$.

Focusing on pauses and silences, the quasi-silence-based segmentation reduces false change point detection, leveraging the natural tendency to pause before speaking. This method improves accuracy without increasing missed detections as opposed to the classical method [7, 8]. However, calculating $BIC$ requires three covariance matrices per sub-window pair (refer Eqns. 1-3), which is computationally heavy with complexity of $O(Nn^2)$, where '$N$' is the number of frames in a window and '$n$' is the number of *MFCC* features [16]. To address this, *Hotelling's $t^2$- Statistic* [21] is proposed as an alternative statistical divergence measure.

### C. Hotelling's T-Squared Statistic

Hotelling's $t^2$-statistic [21] generalizes the student $t$-statistic, that calculates the ratio of estimated error to standard error, in a 12-dimensional feature vector for this work. It uses consecutive *MFCC* vectors in a window to identify speaker change points. The *$t^2$-statistic* is given by:

$$T^2 = \frac{N_x N_y}{N_S}\left(\overline{\mu_x} - \overline{\mu_y}\right)^T \Sigma^{-1}\left(\overline{\mu_x} - \overline{\mu_y}\right) \qquad (4)$$

In other words, the problem can be stated as follows:

$$\max_{\overline{\mu_x}, \overline{\mu_y}, \Sigma} \left[\delta_{x,y}^{\mathrm{T}}(\bar{\mu})\Sigma^{-1}\delta_{\mathrm{x,y}}(\bar{\mu})\right] \qquad (4a)$$

$$\text{s.t.} \quad \delta_{x,y}(\bar{\mu}) = \overline{\mu_x} - \overline{\mu_y}, \quad \forall S = 1, 2, 3, \ldots \quad (4b)$$

$$|\Sigma| \neq 0 \qquad (4c)$$

$$|\Sigma| \succ 0, rank(\Sigma) = \dim(\chi) \qquad (4d)$$

Here, $\overline{\mu_x}$ and $\overline{\mu_y}$ are sample means of sub-windows $X$ and $Y$, and $\Sigma^{-1}$ is the inverse of the covariance matrix of the window $S$. Equation 4 shows that *$t^2$-statistic* requires 1 covariance matrix calculation, making it computationally efficient compared to $BIC$, which requires 3 covariance matrices.

### D. T-Squared Statistic for Segmentation with Quasi-Silence

The following procedure segments around quasi-silences using the *$t^2$-statistic* and is given below.

**Algorithm 2:** Quasi-Silence Based Segmentation using *$t^2$-statistic.*

**Input:** Signal '$S$' and list of quasi-silences '$Qs$'
**Output:** Segment boundaries '$I$'
**for** $j = 0$ to $length(Qs)$ **do**
  $A_W = S(1, N_{aw})$ *// Analysis window around Qs*
  $W = S(1, N_w)$ *// Window of size $N_w$ inside $A_W$*
  $(t_{max}^2, i_{max}) \longleftarrow$ compute *$t^2$-statistic*
  **if** $N_w < N_{aw}$ **then do**
    **if** $t_{max}^2 > 0$ **then do**
      $\Delta BIC_{max} \longleftarrow$ compute $\Delta BIC$ at $i_{max}$
      **if** $\Delta BIC_{max} > 0$ **then do**
        $i_{change} = i_{max}$
      **end if**
      Slide Window by $N_{slid}$
    **else**
      Grow Window by $N_g$
    **end if**
  **end if**
  $I \longleftarrow i_{change}$
**end for**

Unlike Algorithm-1, the *$t^2$-statistic* is computed for the sub-windows to identify quasi-silence locations with high probability of speaker change. At the quasi-silence with the highest *$t^2$-statistic*, $BIC$ determines speaker change existence, making it a hybrid of both. Sliding and growing operations are the same as in Algorithm-1. This hybrid method maintains the benefits of using quasi-silences while reducing redundant $BIC$ calculations. Similar segmentation accuracies are achieved with less computational effort. Clustering these segments further reduces the computational burden on automatic speaker identification (ASI) by avoiding segment-by-segment decisions, which is discussed next.

### E. Greedy Clustering using Bayesian Information Criterion

Segments from the segmentation stage are short, and often resemble auditory noise, making speaker identification computationally redundant and less informative. To address this, segments are clustered based on acoustic feature similarity using $BIC$. Lower $BIC$ values indicate similarity, leading to clustering decisions. This reduces the sample space for speaker identification from many small segments to fewer, larger clusters, streamlining the process and improving computational efficiency for the pre-trained ASI.

Discussion: The selection of Hotelling's *$t^2$-statistic*, $BIC$, and Cepstral Analysis using *MFCC*s is grounded in their distinct technical advantages and their synergistic contributions to speaker identification.

Hotelling's *$t^2$-statistic* is chosen for its robust capability in detecting changes in multivariate distributions, crucial for identifying speaker change points in the high-dimensional 12-D *MFCC* feature vectors. Its computational efficiency is a key advantage, as it requires only one covariance matrix per window compared to three in $BIC$. By evaluating variations in means of multivariate distributions, *$t^2$-statistic* effectively scans audio conversations for speaker changes, making it both simple and rigorous for segmentation.

The hybrid method with $BIC$ complements *$t^2$-statistic* by balancing model complexity and goodness of fit, preventing over-segmentation in unsupervised tasks. $BIC$ integrates a penalty for model parameters, mitigating overfitting and ensuring appropriate model complexity, especially with limited data. This is crucial as the audio samples within a window/sub-window are measured in milliseconds, necessitating a method that can model limited data efficiently.

Note that *MFCC*s is used due to its ability to deliver a robust feature set that accurately represents the speech signal. Furthermore, $BIC$ and the Hotelling's *$t^2$-statistic* are preferred over measures like Kullback-Leibler divergence (*KLD*) for their ability to handle model complexity and avoid segmentation bias. *KLD*, while effective in quantifying dissimilarities, introduce bias due to its asymmetry, therefore making $BIC$ and *$t^2$-statistic* more suitable for speaker segmentation.

## V. Federated Learning for Speaker Identification

Common Automatic Speaker Identification (ASI) systems, like those utilizing Neural Networks (NN) [30, 31],

rely on substantial training data for accurate source prediction in conversational audio signals.

### A. Drawbacks of Centralized Training

Centralized training requires large, unbiased datasets from all speakers, but mobile devices often provide data for only one or a subset of speakers, reducing system accuracy. To overcome this, a distributed diarization framework is proposed, but sharing data across devices raises communication cost concerns. Federated Learning (FL) [19, 20] emerges as a solution for speaker identification system training without sharing data across user devices.

### B. Federated Training

In this collaborative decentralized approach, a *central server or arbitrator* aggregates updates from client devices with identical model architectures (refer Fig. 1). Local updates occur on each client '$i$', and global updates aggregate model parameters [20] or gradients at the central server. Weights are assigned based on a client's contribution, for example number of data samples per client '$n_i$'. The weight assignment '$W_i$' and model aggregation '$\theta_{aggregated}$' for improved and accurate speaker identification are shown below in Eqns. 5 and 6.

$$W_i = \frac{n_i}{\sum_{\forall i} n_i} \tag{5}$$

$$\theta_{aggregated} = \frac{\sum_{i=1}^{m} W_i . \theta_i}{\sum_{i=1}^{m} W_i} \tag{6}$$

### C. Effect of IID and Non-IID Audio Data

The speaker identification performance in diarization systems heavily relies on data uniformity during client updates. *Class imbalance* can lead to *non-IID* data, which may affect the model's performance due to skewed training data [19, 20]. In this work, we employ random client device selection methods for federated aggregation and high initial learning rate for client updates which can reduce the detrimental effect of *non-IID* data in speaker identification.

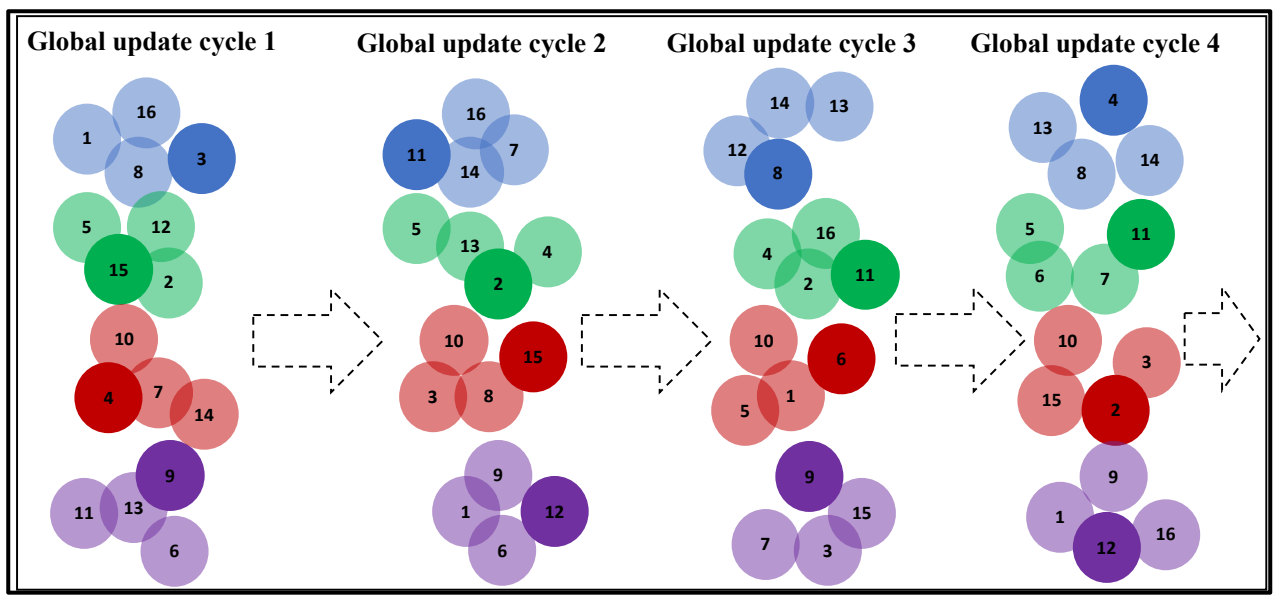

Fig. 2. Random client grouping for global update of Federated Learning

### D. Federated Aggregation for Speaker Identification

In the speaker diarization system, each client device's ASI is trained with data from a single speaker (*non-IID*). The NN uses 12-D *MFCC* feature vectors as input. Hidden layer sizes are empirically chosen, with a *softmax* [38] output layer for one-hot encoded [30] speaker labels. This layer also functions as an embedding layer in the model update phase, elaborated in Section V E.

During training, each client device undergoes local updates following an aggregation method that is the key in FL algorithms. Some of the popular aggregation methods are Federated Averaging (*FedAvg*) [43], Federated Stochastic Gradient Descent (*FedSGD*) [44], Adaptive Federated Optimization (*FedOpt*) [42], Federated Proximal (*FedProx*) [45], Federated Averaging with Momentum (*FedAvgM*) [45] etc. Here, the standard *FedAvg* algorithm [19] is used due to its simplicity in implementation and effectiveness. This is more effective when the sample data at each client device is drawn from the same population. However, due to the *non-IID* nature of speaker data at each device, the global update phase diverges from the straightforward *FedAvg* process. Additionally, the participants for model aggregation can't be determined *a proiri* due to the *ad hoc* nature of client devices location and accessibility. For this work, we have adopted a distributed aggregation approach where clients are randomly grouped, maintaining consistent group sizes across the network, with one client per group randomly selected as the central arbitrator (opaque solid colors in Fig. 2) for a global update cycle. Straggler clients and central arbitrators (group leaders) are handled using timeouts within the neighbor discovery polling method. When a client does not report back within a designated timeframe, the group leader moves forward with aggregating the updates from responsive clients. If a group leader becomes a straggler, another client is dynamically chosen as the new leader to maintain seamless operation. Neighbor discovery polling involves small poll and response packets, usually just a few bytes, which significantly reduces communication overhead compared to data packets. This approach minimizes retransmissions and lowers computational demands on devices acting as central arbitrators by dynamically adjusting group composition. To counteract the impact of *non-IID* data on model updates, an initially high learning rate for local updates is employed. This high learning rate, coupled with random client grouping (*RCG*), emulates *IID* data behavior with larger weight gradients. The learning rate is systematically reduced over cycles to ensure learning convergence, mitigating performance degradation associated with *non-IID* data in FL-based models.

### E. Online Update of Federated Learning-based Speaker Diarization System

In a conversational environment, unpredictability can result in random participants, leading to low initial prediction accuracy for speaker identification models. To improve accuracy, *on-the-fly* model updates can be implemented, as shown in Fig. 3. The neural network's output layer on client devices, without *softmax* activation, serves as an embedding layer of size $L_e$, representing a low-dimensional real-valued vector of the speaker's acoustic characteristics.

For online updates, when segments from a cluster (out of $D_C$ clusters) are processed by a trained ASI, the model predicts the speaker's identity and compares the segment's embedding $CD$ with $TD$, the embeddings of training audio data $D_T$. High similarity results in a model update, while low similarity means the system predicts the speaker's identity without updating the model. *Cosine Similarity* ($S_C$) is used to compute this similarity, with a high value prompting model updates.

$$S_C = \frac{1}{D_C D_T} \times \left[ \sum_{k=1}^{D_C} \left\{ \sum_{j=1}^{D_T} \left( \sum_{i=1}^{L_e} \left( TD_i^j . CD_i^k \right) \right) \right\} \right] \tag{7a}$$

This equation sums the product of corresponding elements from the embedding vectors of training data ($TD$) and cluster segment data ($CD$). The Euclidean Norm of the embedding vectors for training and cluster segments ensures normalization, unbiased by the magnitude of the vectors:

$$\left\|TD^j\right\| = \sqrt{\sum_{i=1}^{L_e}\left(TD_i^j\right)^2}\ ;\ \left\|CD^k\right\| = \sqrt{\sum_{i=1}^{L_e}(CD_i^k)^2} \quad (7b)$$

Using the norms, the normalized cosine similarity is given as:

$$S_C = \frac{1}{D_C D_T} \times \left[\sum_{k=1}^{D_C}\left\{\sum_{j=1}^{D_T}\left(\frac{\sum_{i=1}^{L_e}\left(TD_i^j . CD_i^k\right)}{\left(\sqrt{\sum_{i=1}^{L_e}\left(TD_i^j\right)^2}\right).\left(\sqrt{\sum_{i=1}^{L_e}(CD_i^k)^2}\right)}\right)\right\}\right] (7c)$$

This method ensures accurate speaker identification by updating the model only when necessary, based on high similarity in embeddings.

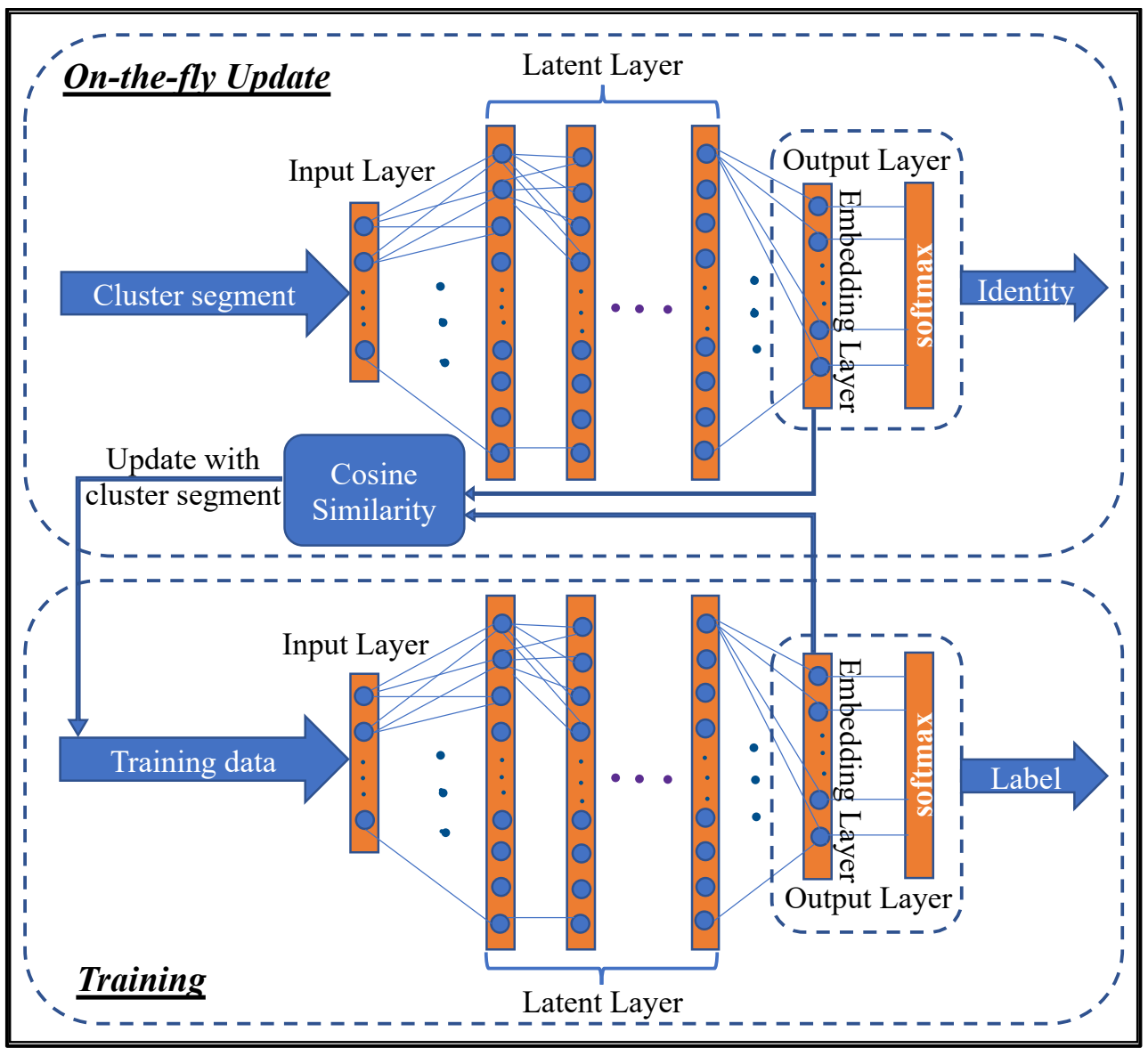

Fig. 3. On-the-fly local update of client model in Federated Learning-based diarization

Note that the aforementioned method is computationally efficient and the foremost reasons contributing to the reduction in computational complexity is via two perspectives. First, the central arbitrator's (server's) complexity is reduced due to the number of communication and computational rounds they are subjected to while aggregating the models. The reduction in communication rounds is ascribed to the subset of client devices participating during each global update. The reduction in computational rounds per device is ascribed to the segmentation state. To be noted that a speech sample is used to train the model after successful identification of a segment. This will depend on the adeptness of two stages viz. the segmentation and the identification stage. Therefore, with unsuccessful segmentation and speaker identification for a speech segment, the server-side may encounter no participation from its respective client group. Second major contributor towards the substantial reduction in computation is ascribed to the *RCG* mechanism, mentioned in Section V C. Here, *RCG* entails arbitrary selection of contributing client per group which results in random selection of a group leader (server) amongst the group. This mechanism divides the computational load across all client devices in a distributed manner, therefore relaxing the computational allocation and burden one server device.

**TABLE I**
EXPERIMENTAL SETUP

| Parameters | Description |
|---|---|
| $\theta_0$ | $\theta_0 \sim \mathcal{N}\left(0, \sqrt{\frac{2}{n_{in}}}\right)$ |
| Client Data Distribution | *Non-IID* data distribution |
| $E$ | 10 (1 $for\ FedSGD$) |
| $\theta_t^k$ or $\Delta L_k(\theta_t)$ | $\theta_t^k$ or $\Delta L_k(\theta_t)$ for *FedSGD* |
| $\theta_{t+1}$ | $\sum_{k=1}^{K}\frac{n_k}{N}\theta_t^k$ or $\theta_t - \eta\sum_{k=1}^{K}\frac{n_k}{N}\Delta L_k(\theta_t)$ or $\theta_t + \beta v_t - \eta\sum_{k=1}^{K}\frac{n_k}{N}\Delta L_k(\theta_t)$ |
| $\mu\|\|\theta - \theta_t\|\|^2$ | $\mu\|\|\theta - \theta_t\|\|^2$ |
| $\mu$ | 0.01 |
| $v_t$ | $v_{t+1} = \beta v_t + (1-\beta)\Delta\theta_t$ |
| $\beta$ | 0.9 |
| $\eta$ | 0.01 (0.9 for *RCG*) |
| $\gamma$ | 0.1 (for *RCG*) |
| $T$ | 200 (or 2000 for *FedSGD*) |
| Straggler Mitigation | Fastest clients for aggregation |
| $f_{stragglers}$ | 0.3 |
| $N$ or $G$ | $N = \sum_{k=1}^{K} n_k$ or $N_g = \sum_{k=1}^{G} n_k$ |

## VI. Experimental Setup and Performance Metrics

### A. Dataset

The study utilized a diverse collection of concise audio conversational datasets with varying speaker change points, ranging from 3 to 20 per conversation. A total of 82 audio files from diverse sources, including Zoom meetings, YouTube content, podcasts, and recordings, ensured comprehensive validation of the proposed diarization system. The deliberate inclusion of diverse sources, databases, and gender representation ensures the broad-spectrum validation of the proposed diarization system.

### B. Implementation

Two Quasi-Silence-Based segmentation methods were employed, differing in sub-window distance calculation within the analysis window. Quasi-silences were determined by thresholding the mean squared energy of a frame at a signal-to-noise ratio of 60 decibels [17, 18], following spectral subtraction. Speaker change points were analyzed within a 1.75-second window surrounding a quasi-silence event, using various window sizes and strides. The diarization system identifies 12 speaker classes, representing participants, with a neural network. Each client device trained independently on its speaker's audio data, with a maximum duration of 20 minutes per session. To be noted that the audio data for training is converted into 12 dimensional acoustic features extracted using *MFCC*. Online updates treated each cluster of audio segments as an epoch. A fixed initial learning rate of 0.9 was adopted for *RCG* federated aggregation, utilizing the Adam optimizer [38]. For centralized training and *non-IID* data training, learning rates decayed over epochs during online model updates. The experiments are conducted for the proposed *FedAvg* with Random Client Grouping and base baseline methods like *FedSGD*, *FedAvg*, *FedProx* and *FedAvgM* for which the parameters are shown in Table I.

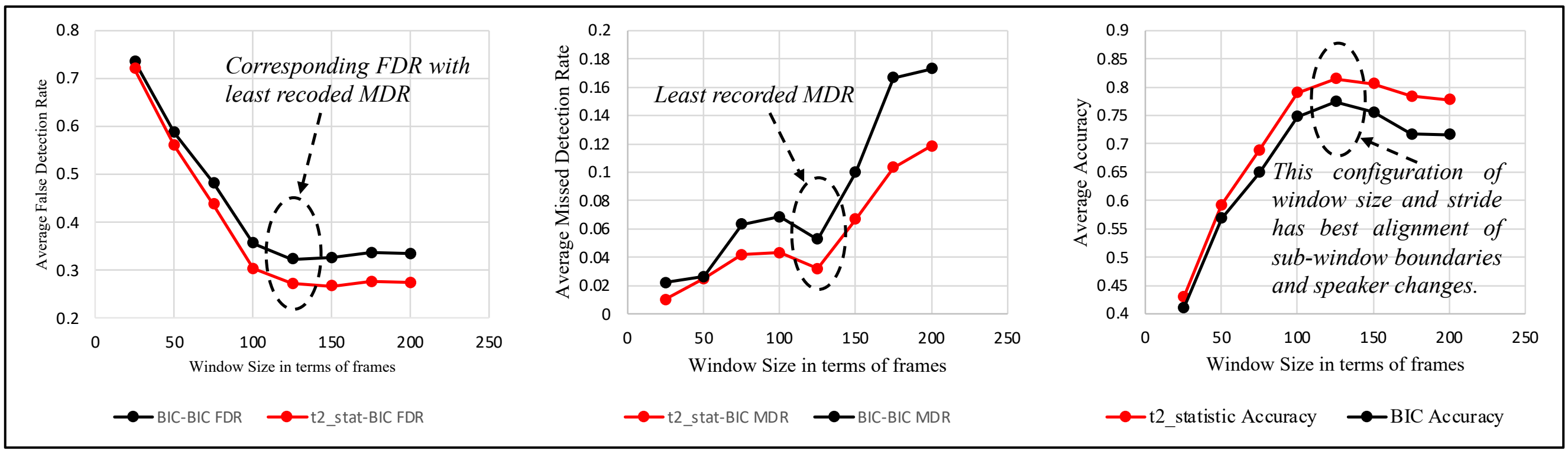

Fig. 4. (Left to right) Comparison of Average FDR, MDR and F-score Accuracy with changing window size respectively.

### A. *Evaluation Metrics*

The performance metrics used to evaluate the system include False Detection Rate (FDR), Missed Detection Rate (MDR), and F-score for speaker change detection. Additionally, purity and coverage metrics are used to assess the system's overall performance. For speaker identification, the evaluation also employs the F-score, considering both the False Acceptance Ratio (FAR) and False Rejection Ratio (FRR). Collectively, these metrics gauge the efficacy of the proposed segmentation model in detecting speaker changes and the precision of speaker identification compared to existing methods. These are mathematically expressed as:

$$FDR = \frac{Detected - (True \cap Detected)}{Detected} \tag{8}$$

$$MDR = \frac{True - (True \cap Detected)}{True} \tag{9}$$

$$F_{seg} = \frac{2*(1-FDR)*(1-MDR)}{2-FDR-MDR} \tag{10}$$

$$Purity = \frac{\sum_{Number\ of\ conversations} |True \cap Detected|}{\sum_{Number\ of\ conversations} Detected} \tag{11}$$

$$Coverage = \frac{\sum_{Number\ of\ conversations} |True \cap Detected|}{\sum_{Number\ of\ conversations} True} \tag{12}$$

$$F_{ID} = \frac{2*(1-FAR)*(1-FRR)}{2-FAR-FRR} \tag{13}$$

## VII. Experimental Results

Performance evaluation using the above indices were done to determine the efficacy of the proposed diarization system. First the segmentation accuracy is reported followed by the speaker identification accuracy, so that the together they can be used as a combinatorial metric for diarization performance.

### B. *Segmentation Accuracy Comparison*

*1) Changing Window Size:* The segmentation accuracy $F_{seg}$ was evaluated using the *f-score* metric, derived from the *FDR* and *MDR* as shown in Equations 8-10. Fig. 4 shows the variations in *FDR*, *MDR*, and $F_{seg}$ with increasing window sizes. The $t^2$*-statistic* hybrid segmentation significantly improved *f-score* accuracy compared to the baseline *BIC*-based method, with both methods using quasi-silences to locate speaker change points.

Accuracy increased by 3% to 8% for window sizes of 100-150, with the 125-window size achieving the lowest *MDR* and highest $F_{seg}$ of 82.5%. This improvement is due to the homogeneity of speech data within each sub-window and heterogeneity across quasi-silences, which enhances accuracy without raising *MDR*. The $t^2$-statistic is effective

**TABLE II**

Comparison of FDR, MDR and F-score Accuracy with respect to different window strides

| Window Size (in frames) | Window Stride | *BIC*-Quasi-Silence-based Segmentation | | | $t^2$-*statitic*-Quasi-Silence-based Segmentation | | |
|---|---|---|---|---|---|---|---|
| | | FDR | MDR | F-SCORE | FDR | MDR | F-SCORE |
| 100 | 0.2 | 0.45978283 | 0.1136186 | 0.65587749 | 0.34626524 | 0.1017986 | 0.75057344 |
| 100 | 0.4 | 0.4793486 | 0.15341241 | 0.62978266 | 0.32607939 | 0.13914241 | 0.75296158 |
| 100 | 0.6 | 0.4082396 | 0.06159738 | 0.70855514 | 0.31668682 | 0.05311738 | 0.788714 |
| 100 | 0.8 | 0.39093677 | 0.05041456 | 0.72338622 | 0.30062781 | 0.03961456 | 0.80204336 |
| 125 | 0.2 | 0.38495157 | 0.0494448 | 0.73062351 | 0.31253264 | 0.02377236 | 0.80375407 |
| 125 | 0.4 | 0.3546287 | 0.04215236 | 0.75514065 | 0.31561428 | 0.0349948 | 0.79384184 |
| 125 | 0.6 | 0.32752207 | 0.05538926 | 0.76921918 | 0.27230291 | 0.03952926 | 0.82570387 |
| 125 | 0.8 | 0.28265795 | 0.08075278 | 0.7888577 | 0.20207007 | 0.07446278 | 0.85128949 |
| 150 | 0.2 | 0.41085073 | 0.09521035 | 0.69850287 | 0.31582108 | 0.08076035 | 0.77773037 |
| 150 | 0.4 | 0.35739511 | 0.09312957 | 0.73627802 | 0.26445985 | 0.08113957 | 0.80964265 |
| 150 | 0.6 | 0.33869876 | 0.12660531 | 0.73557836 | 0.21220185 | 0.11684531 | 0.82563356 |
| 150 | 0.8 | 0.3993331 | 0.56233617 | 0.47472506 | 0.24286487 | 0.54880617 | 0.56431727 |

for modeling distributions with limited samples, which is advantageous for small window sizes (~150 frames).

The empirical choice of a window stride of 25, roughly 20-30% of the window size, ensured optimal segmentation performance. The presence of some silence in both sub-windows had minimal impact due to their comparability.

*2) Changing Window Stride:* Table II compares the performance of two segmentation methods with varying window strides. For window sizes of 100 to 150, strides of 20%, 40%, 60%, and 80% were tested. The $t^2$*-statistic* hybrid segmentation achieved the highest $F_{seg}$ (~85%) with a window size of 125 and an 80% stride, however this also led to a higher $MDR$. The optimal stride range was found to be 20-40% of the window size which achieves comparable performance without significant $MDR$.

The $t^2$*-statistic* hybrid segmentation improved accuracy due to reduced $FDR$ and $MDR$, as shown in Fig. 4 and Table II. This method benefits from considering sub-window mean differences for change point detection, making it suitable for sparse data. It calculates the squared distance between sub-window means, then leverages $BIC$'s model assumption. Additionally, the method focuses on quasi-silences, which aligns sub-window boundaries with speaker change points, and uses a small analysis window to minimize switches, therefore reducing $FDR$ and enhancing $F_{seg}$.

### C. Coverage and Purity Comparison

Fig. 5 compares the coverage and purity between the $BIC$-based and $t^2$*-statistic* hybrid methods (see Eqns. 11 and 12). The hybrid method shows a 3% improvement in coverage and a 5% enhancement in purity at a window size of 125, highlighting its superior change point detection. Note that coverage indicates the method's certainty, while purity measures segmentation precision. The $t^2$*-statistic* hybrid with quasi-silence-based segmentation consistently outperforms the $BIC$-based method, prioritizing coverage to avoid miss detection of critical change points.

### D. Computation Complexity

Fig. 6 shows average $\Delta BIC$ calculations, crucial for window size and stride analysis. The $t^2$*-statistic* hybrid quasi-silence method requires comparatively fewer covariance matrix computations with complexity of $O(Nn^2)$, where '$N$' is the number of frames in a window and '$n$' is the number of *MFCC* features (refer Section IV), therefore reducing computational load.

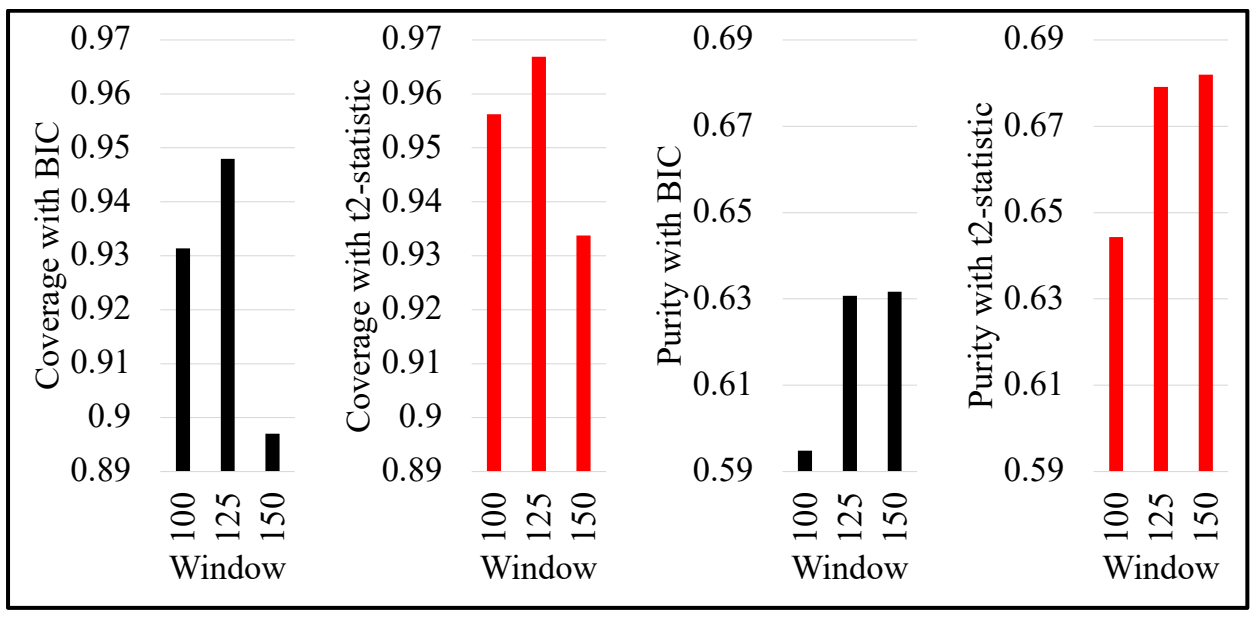

Fig. 5. (Left to Right) Coverage and purity using *BIC* and $t^2$-statistic based segmentation respectively

Overall diarization success relies on accurate segmentation and speaker identification. The performance of the proposed FL-based Speaker Identification System is presented next.

### E. Identification Accuracy Comparison

The speaker identification accuracy has been computed in the following results. These results describe the effect of training a client device with *non-IID* data (only one speaker) and an *IID* approximation by random client grouping as explained in Section V. Fig. 7 shows speaker identification accuracy with classical centralized training, *non-IID* data training and *RCG* for federated training.

Fig. 7 compares the performance of speaker identification stage for diarization system with different training paradigms. The observations are as follows. First, the best identification accuracy is achieved for centralized training. This effect is due to the unbiased data available for training a central device. Since the initial identification accuracy is better than other training paradigms, accuracy increases with a higher slope and then saturates asymptotically. Second observation is that the identification accuracy for the system trained with *non-IID* data shows poor identification accuracy. This is due to the highly biased dataset available to each client device. Since, the initial speaker predictions are erroneous, the online model update (mentioned in Section V E) suffers as well. Third, for *RCG* with 2 clients in each group, the accuracy is better than *non-IID* data training. To be noted that the initial prediction accuracy is low since the initial weight updates are biased use to limited variance in the data samples used for training. However, this system outperforms the system with *non-IID* configuration of training. Fourth, with increase in group size for the *RCG*, the speaker identification accuracy increases. It can be seen that for a group size of 4, the prediction accuracy is close to the accuracy of the centralized training

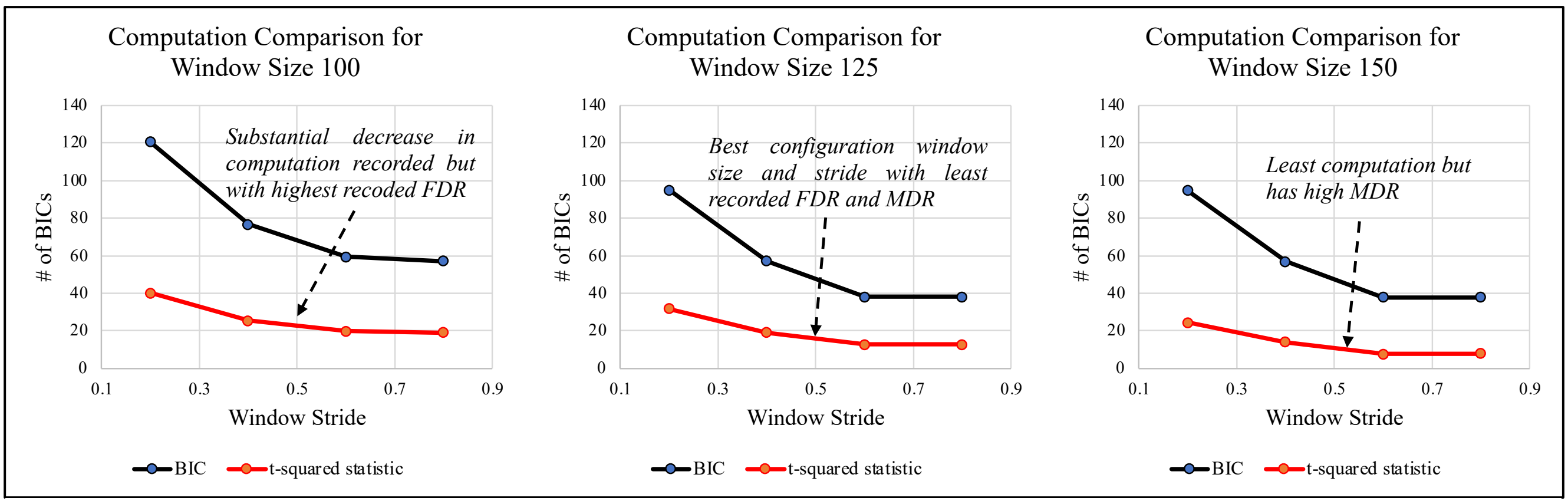

Fig. 6. Average number of *BIC* calculations for three window sizes with changing strides

TABLE III

COMPARISON OF FEGAVG_RCG WITH OTHER STANDARD METHODS

| Method | Accuracy (%) | Total Training Time (hours) | Number of Communication Rounds (T) | Local Computation Steps (E) | Model Update Parallelization | Speaker Label Required for Model Update | Remarks |
|---|---|---|---|---|---|---|---|
| **Centralized DNN** | 89 | 1 | N/A | N/A | N/A | Yes | Baseline centralized training |
| ***FedAvg* [43]** | 87 | 3 | 270 | 10 | N/A | Yes | Stable convergence |
| ***FedSGD*** | 85 | 5 | 3100 | 1 | N/A | Yes | Frequent communication, more overhead |
| ***FedProx* [45]** | 87.5 | 3.5 | 270 | 10 | N/A | Yes | Handles heterogeneity well |
| ***FedAvgM*** | 88.5 | 2.5 | 220 | 10 | N/A | Yes | Faster convergence with momentum |
| ***FedAvg_RCG*** | 86.6 | 6 | 290 | 10 | Yes | No | Handles heterogeneity and converges without explicit computational load clients |

configuration. A detailed comparison of state-of-the-art methods for distributed speaker identification is provided in Table III, demonstrating the efficacy of *FedAvg_RCG* in handling *non-IID* audio data. Table III also shows that *FedAvg_RCG* supports model update parallelization, enabling multiple updates within a single round, unlike its competitors. From the perspective of a client device, data is transmitted only once per communication round, but models are updated simultaneously by multiple group leaders. This approach achieves comparable performance with significantly reduced communication and computation load on the group leaders compared to a static central coordinator. Finally, it can be observed from Fig. 7 that the learning speed is different for all the configurations. The learning latency decreases as the training paradigm moves from non-*IID* to *IID* data. This means that for the non-*IID* data training, the increase in accuracy is sluggish. Whereas the centrally trained system learnt promptly (slope is high).

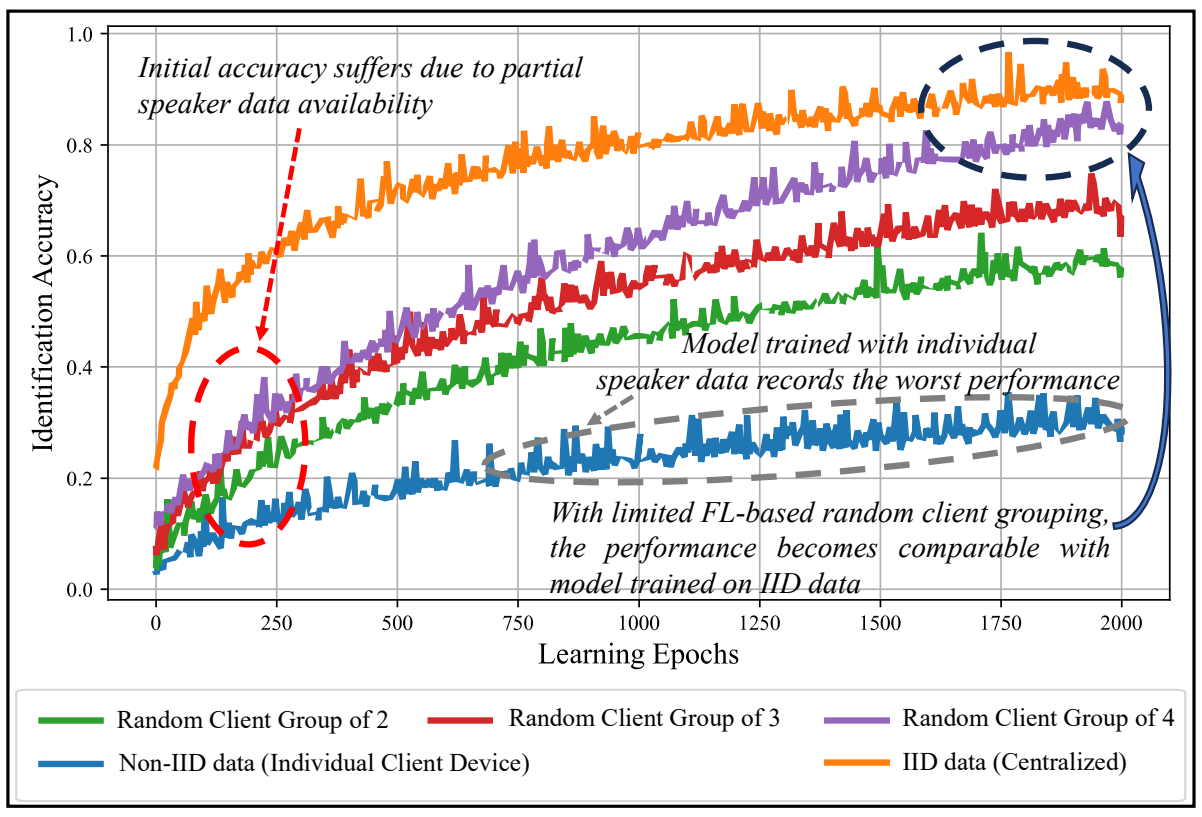

Fig. 7. Identification Performances for different training configurations

<u>*Discussion*</u>: Understanding random client grouping for training and its impact on weight updates emulating *IID* behavior is crucial. In *IID* datasets, samples are fed randomly, meaning consecutive training samples have varying class labels. This randomness leads to initial weight updates to be random across the weight space $\mathbb{W}$. Conversely, *non-IID* data biases weight updates, limiting them to specific regions in $\mathbb{W}$. Here, the weight space represents the hyperspace of model parameters. Using federated averaging with *RCG*, subsets of client devices update weights with a high learning rate, moving in the direction of the raw gradient. When aggregated at the group's central arbitrator, these updates appear stochastic due to changing client groups each training epoch, mimicking *IID* behavior. This stochasticity helps avoid local minima, a common issue with biased *non-IID* updates. However, it may delay learning convergence. As the identification stage trains online and identifies participating speakers, the diarization system's performance improves.

<u>*Challenges and Mitigation Strategies*</u>: Designing the FL-based speaker diarization system posed several potential challenges. One challenge is the heterogeneity of data across devices, each recording different environments and speakers. This work addresses this with robust client device grouping for federated model aggregation, ensuring diverse audio data contributes to the model.

Another challenge is under-segmentation and over-segmentation. The proposed method uses $t^2$-*statistic* between consecutive speech frames to tackle under-segmentation, and $BIC$ at potential speaker change points to address over-segmentation. Considering embedded devices' computational limitations, lightweight algorithms like quasi-silences and $t^2$-*statistic* have been employed for speaker segmentation, minimizing computational load.

## VIII. CONCLUSION

The proposed Unsupervised Federated Learning based Speaker Diarization system provides an end-to-end design for an IoT-style audio network. It tackles the design problem from three perspectives, namely conversation segmentation, clustering, and speaker identification. It proposes an unsupervised Hotelling's $t^2$-statistic-quasi-silence-based segmentation method which improves segmentation accuracy by finding dissimilarity between conversation segments. Additionally, this method limits the computational rigor in two ways; (a) limiting the speaker change search around quasi-silences, and (b) limiting the number of *BIC* calculation by employing $t^2$-statistic. Further, the proposed method achieves unsupervised segment clustering using Bayesian Information Criterion to find similarity between segments. Finally, identity of the clusters is deciphered by FL based Speaker Identification model trained at each IoT device. These models use a random client grouping technique to train each speaker identification model which can emulate *IID* speaker data distribution. A cosine similarity based online update mechanism is proposed for the FL based speaker Identification model. It is shown that the proposed segmentation methods can achieve segmentation accuracy of nearly 85% without external supervision. This method reports a purity of approximately 97%, which means that it seldom misses a speaker change. The unsupervised online update mechanism in tandem with random client grouping based FL model achieves a speaker identification accuracy of ≈ 90%. This is very close to centralized training with *IID* speaker data distribution. Future works will include analysis of overlapped speech during conversations.